\documentclass[preprint]{elsarticle}

\usepackage{amsmath}
\usepackage{amssymb}

\journal{J. Crystal Growth}

\begin{document}

\begin{frontmatter}

\title{Investigation of the Nd$_2$O$_3$--Lu$_2$O$_3$--Sc$_2$O$_3$ phase diagram for the preparation of perovskite-type mixed crystals NdLu$_{1-x}$Sc$_x$O$_3$}

\author[mymainaddress,mysecondaryaddress]{Tamino Hirsch\fnref{myfootnote1,myfootnote2}\corref{mycorrespondingauthor}}
\fntext[myfootnote1]{This work is based on a thesis submitted by TH to the Institute for Mineralogy, TU Bergakademie Freiberg in partial fulfillment of the requirements for the degree of Master of Science.}
\fntext[myfootnote2]{Current address: Fakulty VI, Technical University Berlin, Gustav-Meyer-Allee 25, 13355 Berlin, Germany.}
\ead{t.hirsch@tu-berlin.de}
\cortext[mycorrespondingauthor]{Corresponding author}

\author[mymainaddress]{Christo Guguschev}

\author[mymainaddress]{Albert Kwasniewski}

\author[mymainaddress]{Steffen Ganschow}

\author[mymainaddress]{Detlef Klimm}

\address[mymainaddress]{Leibniz-Institut f\"ur Kristallz\"uchtung, Max-Born-Str. 2, 12489 Berlin, Germany}
\address[mysecondaryaddress]{Institut f\"ur Mineralogie, TU Bergakademie Freiberg, Brennhausgasse 14, 09596 Freiberg, Germany}

\begin{abstract}
Based on differential thermal analysis (DTA) and X-ray powder diffraction (XRD), a description of the system Nd$_2$O$_3$--Lu$_2$O$_3$--Sc$_2$O$_3$ was obtained by thermodynamic assessment. Four fields of primary crystallization could be identified; from melt compositions close to the Lu$_2$O$_3$--Sc$_2$O$_3$ edge the rare-earth oxide C-phase crystallizes first, which is stable down to room temperature. From Nd$_2$O$_3$ rich melts the X-phase forms, which is stable only at high temperatures. An additional field, where the alternative high-temperature phase H solidifies as primary product touches the Nd$_2$O$_3$--Lu$_2$O$_3$ edge of the concentration triangle. From melts close to the composition NdScO$_3$, the P-phase (perovskite) can be crystallized and mixed crystals with second end member NdLuO$_3$ have been grown from the melt. Crystals of this mixed perovskite were grown by the micro-pulling-down and Czochralski methods.
\end{abstract}


\end{frontmatter}


\section{Introduction}

Rare-earth scandates with orthorhombically distorted perovskite structure (P-REScO$_3$, RE = rare-earth element; in this case Pr--Dy) have come into the focus of interest as substrates for the epitaxial deposition of many functional oxides, and especially for strain engineering of perovskitic layers \cite{Schlom14,Uecker08}. Their pseudocubic lattice parameters $a_\mathrm{pc}\approx4$\,\AA\ can be adjusted by an appropriate choice of the RE element occupying the larger cation site of the perovskite structure. Furthermore, mixed crystals between two rare-earth scandates can be grown and allow fine-tuning of $a_\mathrm{pc}$ \cite{Uecker13,Berndt76,Uecker17,Sayed15,Guguschev18}. Crystals of high quality obtained by the Czochralski technique retain their perovskitic crystal structure during cooling from the melting point down to room temperature.

Perovskitic rare-earth lutetates (P-RELuO$_3$) show even larger lattice parameters than the scandates \cite{Schlom14,Uecker17,Coutures76}, but most of them cannot be prepared from the melt because the perovskite phase is not stable up to the melting point \cite{Coutures80}. The establishment of solid solutions between scandates and lutetates seems to be a feasible way towards substrate materials with larger lattice parameters than that of scandates. This was demonstrated recently with mixed crystals of LaScO$_3$ and LaLuO$_3$ \cite{Uecker17}, which provide a potential substrate for barium stannate (BaSnO$_3$) epitaxy. However, lattice constants that became accessible with LaLu$_{1-x}$Sc$_x$O$_3$ mixed crystals are large (range $4.13\lesssim a_\mathrm{pc}\lesssim4.18$\,\AA) and a gap remains towards PrScO$_3$ ($a_\mathrm{pc}=4.02$\,\AA) which has the largest lattice constant of the rare-earth scandate single crystals that are available so far. One reason for this large gap is the very high melting point $T_\mathrm{f}\approx2660$\,K \cite{Badie78} of LaScO$_3$ which impedes the Czochralski growth of scandium-rich LaLu$_{1-x}$Sc$_x$O$_3$ mixed crystals with $x\gtrsim0.4$. The melting points drop from LaScO$_3$ to the REScO$_3$ phases with smaller RE$^{3+}$ ions, but remain still high for Ce$^{3+}$ and Pr$^{3+}$ \cite{Gesing09}; accordingly these two elements were excluded from the current study.

The next scandate NdScO$_3$ is a well known perovskite substrate \cite{Uecker08}. NdLuO$_3$ is controversially reported to be either a perovskite \cite{Porotnikov80a,Porotnikov83,Coutures74,Berndt75} --- or to form other crystal structures \cite{Schneider60a,Bharathy09,Ito01}. A detailed evaluation of the system NdLuO$_3$ was performed recently \cite{Hirsch17}. Goldschmidt's tolerance factor $t$ indicates that NdLuO$_3$ is just outside the limits of perovskite stability, and should not be a perovskite under equilibrium conditions. Nevertheless, NdLu$_{1-x}$Sc$_x$O$_3$ mixed crystals can be expected to be perovskites beyond some critical scandium concentration $x$. The aim of this paper is to gain some insight into the pseudoternary system Nd$_2$O$_3$--Lu$_2$O$_3$--Sc$_2$O$_3$, with focus on the NdLuO$_3$--NdScO$_3$ section therein --- and to evaluate possibilities for the growth of NdLu$_{1-x}$Sc$_x$O$_3$ mixed crystals. 

Generally, the oxide of a specific 3-valent rare-earth element can occur in one of the following five different solid modifications: the low temperature modifications C (cubic bixbyite structure, $Ia\bar{3}$), B (monoclinic, $C2/m$) and A (hexagonal) and the high temperature modifications H (also hexagonal, $P6_3/mmc$) and X (cubic, $I\bar{3}m$) \cite{Konings14,Adachi98}, see also Tab.~\ref{Gex-meins}. Solid solutions RE$'_{2-x}$RE$''_x$O$_3$ of these modifications (with RE$'$, RE$''$ on the same crystallographic positions) are formed if two rare-earth oxides can be mixed. In contrast, an intermediate perovskite compound RE$'$RE$''$O$_3$ (with RE$'$, RE$''$ on different crystallographic positions) is formed only if the component rare-earth oxides are present nearly in a molar ratio 1:1, and if the difference between the RE$'^{3+}$ and RE$''^{3+}$ ionic radii (and hence Goldschmidt's tolerance factor) is sufficiently large \cite{Badie78,Coutures74,Schneider60a,Coutures76a,Badie70}.

\section{Experimental}

In a previous publication, it was shown that the P-NdScO$_3$ phase is slightly deficient in neodymium and should rather be written as Nd$_{0.967}$ScO$_{2.951}$ \cite{Uecker08}. For the sake of simplicity, ``NdScO$_3$'' is written instead throughout this paper, but the congruently melting ratio of Nd:Sc = 0.967:1.000 was actually used for the experiments. All samples of this study were prepared from powder mixtures on the NdLuO$_3$--NdScO$_3$ isopleth section of the Nd$_2$O$_3$--Lu$_2$O$_3$--Sc$_2$O$_3$ system that corresponds to a horizontal line in the concentration triangle in Fig.~\ref{fig:liquidus} for the idealized NdScO$_3$ (used in thermodynamic assessment) or is close to this horizontal line for Nd$_{0.967}$ScO$_{2.951}$ (used in most experiments). Powders of Nd$_2$O$_3$, Lu$_2$O$_3$ and Sc$_2$O$_3$ with at least 99.99\% purity were used for the samples.

Data for the Nd$_2$O$_3$--Lu$_2$O$_3$ edge were combined out of recent data and the study \cite{Hirsch17}. Thermal analysis of 24 samples ranging from the composition NdLuO$_3$ ($x=0$) to NdScO$_3$ (see previous paragraph, $x=1$) up to the liquidus temperatures was performed with a NETZSCH STA429. For these measurements covered tungsten crucibles with approximately 30--40\,mg sample powder were heated twice in static He atmosphere (99.9999\% purity) at a rate of 15\,K/min above the corresponding liquidus temperature, typically the maximum temperature was 2473\,K. During the first heating the sample was homogenized by melting, then cooled to 773\,K, and subsequently heated again above the liquidus. Mainly DTA curves from these second heating segments were used for the construction of the phase diagram section NdLuO$_3$--NdScO$_3$. The W/Re thermocouples that were used for the detection of DTA signals suffer from degradation due to the preferred evaporation of rhenium \cite{Burns89}. To correct the resulting shift of their emf($T$) function, calibration measurements with pure Al$_2$O$_3$ ($T_\mathrm{f}=2327$\,K \cite{FactSage7_1}) were performed from time to time.

For X-ray powder diffraction studies, four NdLu$_{1-x}$Sc$_x$O$_3$ mixtures with $x=0.1; 0.2; 0.4; 0.6$ were prepared from the oxide powders and were annealed in muffle furnaces in air at 1873\,K for 60 hours. X-ray diffraction (XRD) was performed with a GE Inspection Technologies XRD3003 TT. This instrument uses the Bragg Brentano geometry, a fixed aperture, CuK$\alpha$ radiation (40\,mA, 40\,kV) and a scintillation counter. The range $10^\circ\leq2\Theta\leq90^\circ$ was measured with a step width of $0.01^\circ$ and a detection time per step of 4\,s. All diffractograms shown in this paper display raw data, and lattice parameters of the perovskite phase were calculated from 15 to 16 perovskite peaks that were indexed as given in the literature. Positions of peaks could be determined with software fityk 0.9.8 \cite{Wojdyr10}. From these positions, a generic algorithm in Matlab\textsuperscript{\textregistered} calculated the lattice parameters.

The specific heat capacity $c_p(T)$ of P-NdScO$_3$ powder was measured by heat flux differential scanning calorimetry (DSC) with a NETZSCH STA449C. Three DSC measurements with linear heating segments were performed: first with an empty crucible as baseline, second with Al$_2$O$_3$ powder as reference and third with the sample to analyze. $c_p(T)$ was then calculated by comparison of the obtained curves using the ratio method of NETZSCH Proteus analysis software. All these DSC measurements were carried out in covered Pt crucibles in flowing argon. To ensure reproducibility, four subsequent heating runs with 20\,K/min from 313\,K to 1473\,K were performed during every measurement, and the average of runs 2--4 was used for interpretation.

Crystal growth with the micro-pulling-down method ($\mu$-PD) was performed in a device from N\"urmont Installations GmbH \& Co. KG (Germany). The crystal fibers were pulled from an iridium crucible inside flowing nitrogen atmosphere (99.999\% purity). The starting material had the composition NdLu$_{0.15}$Sc$_{0.85}$O$_3$. For the preparation of the starting materials for Czochralski growth, dried powders of Nd$_2$O$_3$, Sc$_2$O$_3$ and Lu$_2$O$_3$ with purities of 99.99\% (4N) to 99.999\% (5N) were used. The powders were weighed and mixed with the following composition: 60\,mol\% NdScO$_3$ and 40\,mol\% NdLuO$_3$. Subsequently, to optimize the crucible filling process, cylindrical bars were made by cold isostatic pressing at 0.2\,GPa. The Czochralski crystal growth experiment was performed using a conventional RF-heated Czochralski set-up equipped with a crystal balance and automatic diameter control. The atmosphere during growth was 5N argon under ambient pressure. Iridium crucibles (inner diameter: 38\,mm, height: 40\,mm) embedded in ZrO$_2$ and Al$_2$O$_3$ insulation were used. An actively heated iridium afterheater was placed on top of the crucible. A growth rate of 0.7\,mm\,h$^{-1}$ and a rotation rate of 10\,rpm were used. Due to the lack of a seed crystal, an iridium rod was used instead. 

The chemical compositions of three Czochralski crystal pieces and residual material were measured with an IRIS Intrepid HR Duo (Thermo Elemental, U.S.A.) ICP-OES spectrometer. For that purpose, 5--7\,mg material were dissolved in 50\,ml of a solution of 3\,ml HNO$_3$ and 1\,ml H$_2$O$_2$ in water for 20 minutes at $220^{\,\circ}$C. These determinations were repeated 3 times. Besides, parts of the crystals were characterized by XRD. Lattice parameters of one piece were determined with the same Matlab\textsuperscript{\textregistered} algorithm as mentioned earlier.

\section{Results}

\begin{figure}[ht]
\includegraphics[width=0.80\textwidth]{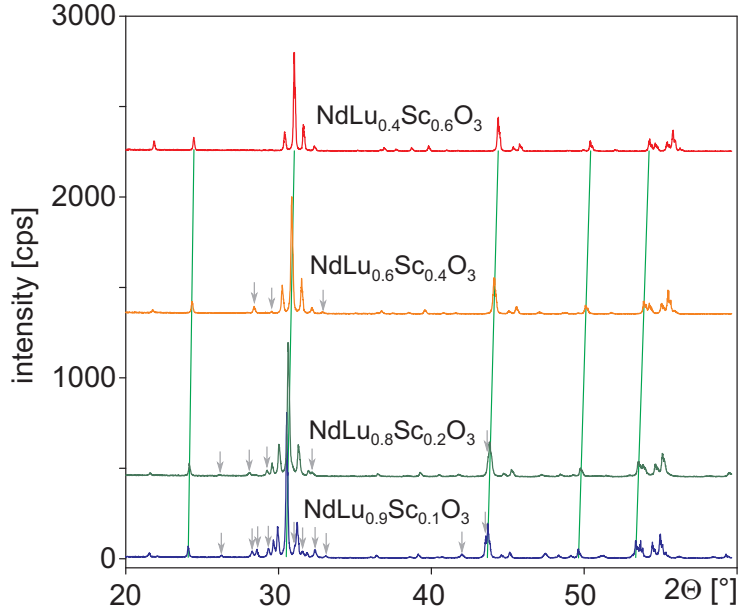}
\caption{XRD pattern of annealed NdLu$_{1-x}$Sc$_x$O$_3$ powders. Only peaks with arrow belong to non-perovskite phases. The three upper curves are parallel shifted. Green lines depict the perovskite peak shift resulting from changing chemical composition.}
\label{fig:x-ray}
\end{figure}

A first experiment was performed to check the stability of the perovskite phase in the subsolidus region. Four NdLu$_{1-x}$Sc$_x$O$_3$ powder mixtures with NdScO$_3$ concentrations that are given in Fig.~\ref{fig:x-ray} were annealed at 1873\,K for 60 hours. The subsequent X-ray powder diffraction analysis revealed even for a sample with 40\% NdLuO$_3$ (top diffractogram in Fig.~\ref{fig:x-ray}) exclusively perovskite peaks. For higher lutetate concentrations additional peaks indicate the coexistence of perovskite with other phases, but even for the $x=0.1$ sample (bottom curve) perovskite is the major phase.

\begin{figure}[ht]
\includegraphics[width=0.7\textwidth]{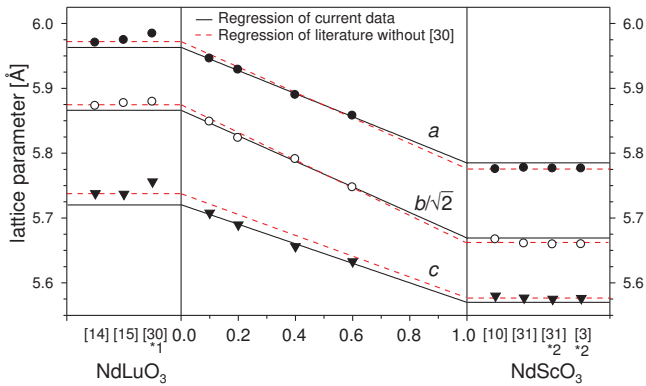}
\caption{Lattice parameters from literature (left and right panels, references at the bottom) for the perovskite phases of NdLuO$_3$ and NdScO$_3$ compared with current data for solid solutions between these endmembers. Division $b/\sqrt{2}$ is for better visibility. *1 material produced by high pressure *2 not from powder but single crystal XRD (Ref. [31] reports single crystal as well as powder data).}
\label{fig:Vegard}
\end{figure}

Especially for larger $2\Theta$ values it is evident that the position of X-ray peaks shifts to smaller values for higher lutetate concentrations. This trend is expected because the Shannon radius of Lu$^{3+}$ (100.1\,pm) is significantly larger than the radius of Sc$^{3+}$ (88.5\,pm) \cite{Shannon76}. Fig.~\ref{fig:Vegard} shows that the lattice parameters $a,b,c$ depend almost linearly on $x$ and follow Vegard's rule \cite{Vegard21}. Comparable behavior was found in the systems CeScO$_3$--PrScO$_3$ \cite{Sayed15} and LaScO$_3$--LaLuO$_3$ \cite{Uecker17,Uecker15}.

These recent results are compared with the literature data of the endmembers (P-NdScO$_3$ and questionable P-NdLuO$_3$). Additionally, Fig.~\ref{fig:Vegard} shows regressions of current data (solid lines) and compares them to a linear (Vegard) behavior between literature data for the endmembers. Lattice parameters for NdLuO$_3$ in the literature are somewhat higher than expected from the recent data, which holds especially for $c$. For NdScO$_3$ the discrepancies are less pronounced.

The coexistence of one phase with variable composition (degree of freedom $F=1$) with at least one other phase (number of phases $P=2$) under isobar ($p=\mathrm{const.}=1$\,bar) and isothermal ($T=\mathrm{const.}=1873$\,K) conditions is not in agreement with Gibbs phase rule for a system with two components ($C=2$), because $P+F=C$ should hold. This means either one phase with variable composition, or two phases with fixed composition are allowed in equilibrium. It can be assumed that either equilibrium could not be obtained during the annealing period, or that partial decomposition of the equilibrium state at 1873\,K occurred during cooling.

Peak shift in the four samples is proportional to the chemical composition of overall system. The absolute peak intensity of the perovskite phase as well as the ratio between perovskite and non-perovskite peaks are approximately constant for all samples $x\leq0.4$. Hence, perovskite seems to be the only equilibrium phase in the system NdLuO$_3$--NdScO$_3$ up to 1873\,K. In the sample with $x=0.1$ non-perovskite peaks are still weak and the change in perovskite lattice parameters is continued, but the composition of that sample could be close to the boundary to a multi-phase field.

Phase compositions at higher temperatures were examined by DTA. A compilation of thermal effects on the NdLuO$_3$--NdScO$_3$ section is given later in Fig.~\ref{fig:NdLuO$_3$-NdScO$_3$-DTA}. All thermal signals during heating were endothermic, and every sample was completely molten during the DTA experiment. On the NdScO$_3$ side DTA curves show less thermal effects. On the NdLuO$_3$ side in contrast, more and weaker peaks down to lower temperatures occur which is obviously related to the instability of the perovskite (P-) phase and the equilibria between B- and C-phase that occur instead \cite{Hirsch17}. The experimental overview in Fig.~\ref{fig:NdLuO$_3$-NdScO$_3$-DTA} should not be considered to be a binary phase diagram, because such binary diagrams can be constructed only between stable phases showing congruent melting. For a better understanding of phase relations on this isopleth section, it must be discussed by using the concentration triangle Nd$_2$O$_3$--Lu$_2$O$_3$--Sc$_2$O$_3$ instead.

\section{Thermodynamic assessment and discussion}

Zinkevich \cite{Zinkevich07} gave a comprehensive overview on thermodynamic data of the RE$_2$O$_3$ compounds (enthalpies of formation $\Delta H_\mathrm{f}$, entropies $S$, transition temperatures $T_\mathrm{t}$, temperature dependent heat capacities $c_p(T)$) in the five solid modifications (C, B, A, H, X) and in the melt (L), which are the base of the work presented here. The $c_p(T)$ of all phases of one specific oxide were assumed identical for reasons of consistency, and only heats of transformation $\Delta H_\mathrm{t}$ were taken into account.

\begin{figure}[ht]
\includegraphics[width=0.6\textwidth]{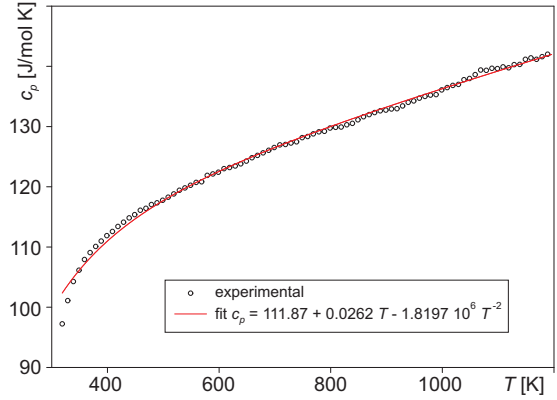}
\caption{Experimental data and fit for the heat capacity of P-NdScO$_3$ (only every $10^\mathrm{th}$ data point is shown).} 
\label{fig:cp-regression}
\end{figure}

For P-NdScO$_3$ $\Delta H_\mathrm{f}=-1\,876\,212$\;J/mol was estimated, based on equation (2) of Qi et al. \cite{Qi15} and the tolerance factor $t$ from ionic radii of Shannon \cite{Shannon76}. A function $c_p(T)$ for P-NdScO$_3$ was published recently by Uecker et al. \cite{Uecker13}. Nevertheless, P-NdScO$_3$ powder was prepared also for the current study via solid state reaction to repeat the previous measurement. Phase purity of the perovskite powder was confirmed by XRD, and $c_p(T)$ is shown in Fig.~\ref{fig:cp-regression} together with an exponential fit. This data seem to be more reliable than the previous data \cite{Uecker13}: firstly they are closer to the average curves of A-Nd$_2$O$_3$ and C-Sc$_2$O$_3$ based on functions given by Zinkevich \cite{Klimm15,Zinkevich07}; and secondly, they show less scatter. Entropy and melting temperature of P-NdScO$_3$ were not input parameters but were optimized by the thermodynamic assessment: for P-NdScO$_3$ $S=121.7$\,J\,mol$^{-1}$\,K$^{-1}$ and for P-NdLuO$_3$ $S=133.9$\,J\,mol$^{-1}$\,K$^{-1}$ was found by this calculation. It should be noted that the formation enthalpy $\Delta H_\mathrm{f,ox}$ for P-NdLuO$_3$ from the component oxides A-Nd$_2$O$_3$ and C-Lu$_2$O$_3$ which can be derived from eq. (2) in \cite{Qi15} is indeed small in absolute value. Consequently this phase is thermodynamically unfavored against a mixture of the oxides. Function $c_p(T)$ of P-NdLuO$_3$ was estimated as the mean of A-Nd$_2$O$_3$ and C-Lu$_2$O$_3$ of Zinkevich \cite{Zinkevich07}.

\begin{figure}[ht]
a) \includegraphics[width=0.46\textwidth]{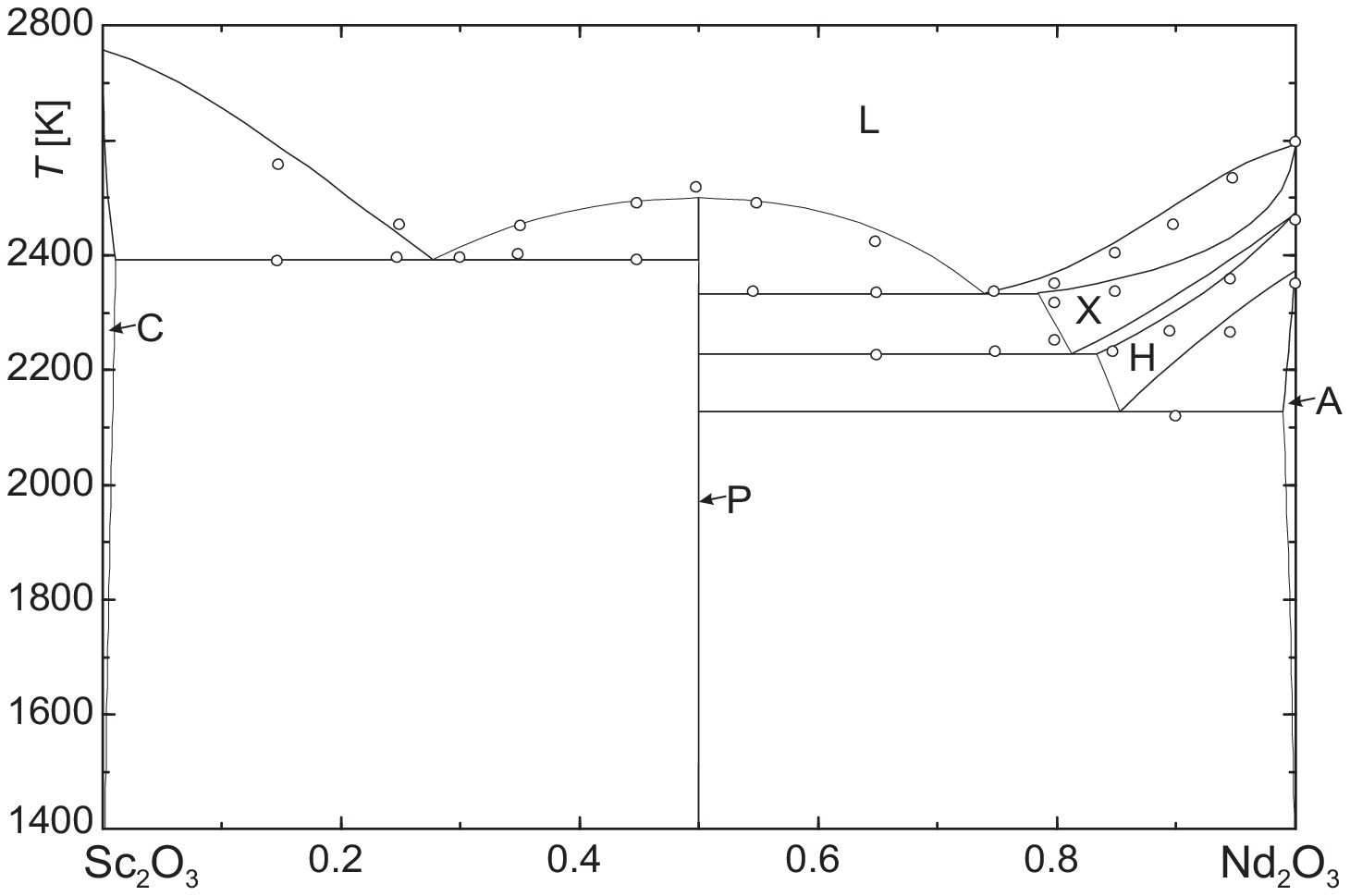} 
b) \includegraphics[width=0.46\textwidth]{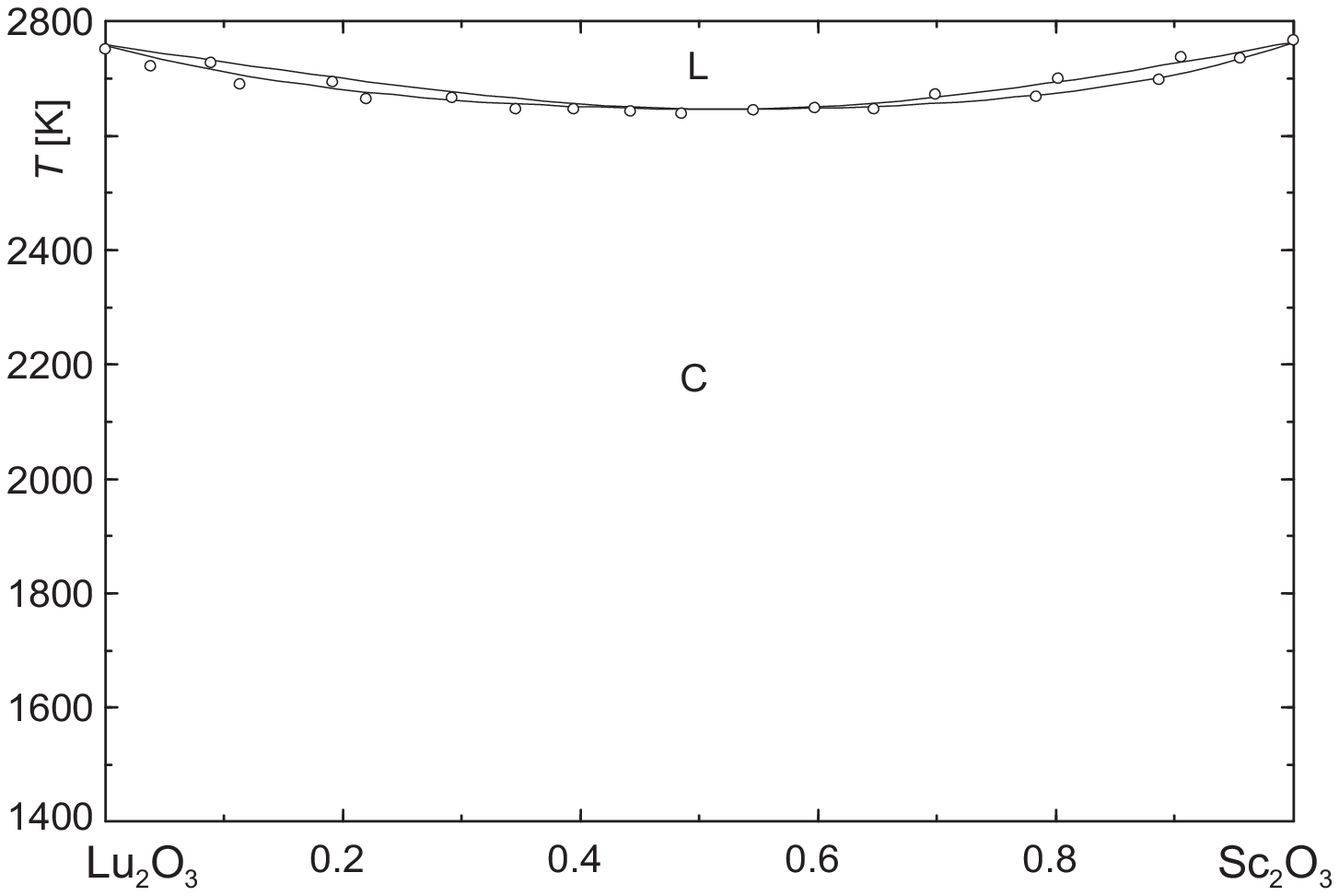}
\caption{Experimental results of Badie \cite{Badie78} (points) for the systems Nd$_2$O$_3$--Sc$_2$O$_3$ and Lu$_2$O$_3$--Sc$_2$O$_3$ together with a FactSage \cite{FactSage7_1} thermodynamic assessment (lines). Labels for 1-phase fields.}
\label{fig:2PDs}
\end{figure}

The assessment of the pseudoternary system Nd$_2$O$_3$--Lu$_2$O$_3$--Sc$_2$O$_3$ was started with models for the three edge systems. Nd$_2$O$_3$--Sc$_2$O$_3$ and Lu$_2$O$_3$--Sc$_2$O$_3$ were experimentally examined already by Badie \cite{Badie78}, and his interpretation is in almost perfect agreement with the model that is presented here (Fig.~\ref{fig:2PDs}). For the mixed phases with solution species $\alpha$ and $\beta$, excess enthalpy functions $G_\mathrm{ex}(x)$ of Redlich-Kister type
\begin{equation}
G_\mathrm{ex}^{\alpha,\beta} = {^0}L x_{\alpha} x_{\beta} + {^1}L x_{\alpha} x_{\beta} (x_{\alpha}-x_{\beta}) + {^2}Lx_{\alpha}x_{\beta}(x_{\alpha}-x_{\beta})^2
\label{eq:Gex-bi}
\end{equation}
with interaction parameters $^iL$ were inserted \cite{Redlich48}. Because most equilibria were considered for a comparably small temperature range at very high $T$, it was not necessary to take an additional temperature dependency $G_\mathrm{ex}(T)$ into account. The excess enthalpy functions of ternary mixtures are defined similarly as
\begin{equation}
G_\mathrm{ex}^{\alpha,\beta,\gamma} ={^0}L x_\alpha x_\beta x_\gamma
\label{eq:Gex-ter}
\end{equation}
with one parameter $^0L$ only. The $^iL$ of the five rare-earth oxide structures and the melt L are summarized in Tab. \ref{Gex-meins}. For the P-phase (P-NdLuO$_3$--P-NdScO$_3$) one finds ${^0}L=34\,000$.

\begin{table}[ht]
\caption{Interaction parameters $^iL$ of binary (eq.~\ref{eq:Gex-bi}) and ternary (eq.~\ref{eq:Gex-ter}) excess enthalpy functions in the system Nd$_2$O$_3$--Lu$_2$O$_3$--Sc$_2$O$_3$.}
\begin{tabular}{cccccccc}
 \hline 
       & Nd$_2$O$_3$-- &  Nd$_{2}$O$_3$--  & Lu$_{2}$O$_3$-- & Nd$_{2}$O$_3$--              & \multicolumn{2}{c}{Nd$_2$O$_3$--} \\
phase  & Lu$_{2}$O$_3$ &  Sc$_{2}$O$_3$    & Sc$_{2}$O$_3$   & Lu$_{2}$O$_3$--Sc$_{2}$O$_3$ & \multicolumn{2}{c}{Sc$_{2}$O$_3$} \\
 \cline{2-7} \rule{0pt}{0.35cm} & \multicolumn{4}{c}{${^0}L$} & ${^1}L$ & ${^2}L$\\  \hline
C   & 27\,000 & 20\,000  & 15\,000 & 20\,000   & --        & -- \\
B   & 10\,500 & 0        & 0       & 20\,000   & --        & -- \\
A   & 20\,900 & -10\,000 & 0       & 0         & --        & -- \\
H   & 17\,200 & -37\,000 & 15\,995 & 20\,000   & --        & -- \\
X   & 12\,600 & -48\,000 & 12\,850 & 0         & --        & -- \\
L   &   0     & -80\,000 & 0       & -30\,000  & -17\,000  & -30\,000 \\
\hline
\end{tabular}
\label{Gex-meins}
\end{table}

The calculated phase diagram Lu$_2$O$_3$--Sc$_2$O$_3$ in Fig.~\ref{fig:2PDs}b) shows no significant deviation from Badie's hand-drawn diagram. For both oxides, the C-phase is stable up to $T_\mathrm{f}$, and they form a solid solution. Almost in the middle of the system an azeotropic point exists. Our model for Nd$_2$O$_3$--Sc$_2$O$_3$ in Fig.~\ref{fig:2PDs}a) is slightly different in comparison to Badie \cite{Badie78} in that sense the P-phase has a fixed 1:1 composition and that the phase width of the A-phase is smaller in the present model. Nevertheless, the phase sequence A$\rightarrow$H$\rightarrow$X$\rightarrow$L of Nd$_2$O$_3$ with remarkable solubility of Sc in Nd$_2$O$_3$ especially in the high-$T$ phases H and X is reproduced well.

\begin{figure}[ht]
\includegraphics[width=0.45\textwidth]{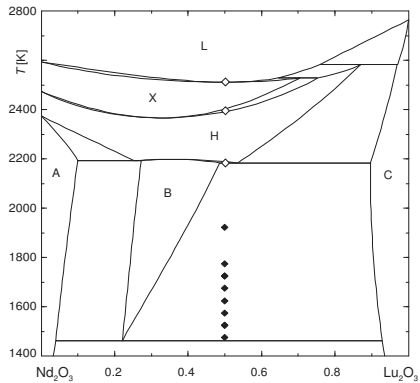}
\caption{DTA (empty diamonds) and XRD (filled diamonds, point at 1923\,K from \cite{Schneider60a}) results for the systems Nd$_2$O$_3$--Lu$_2$O$_3$ together with a FactSage \cite{FactSage7_1} thermodynamic assessment (lines). Labels for 1-phase fields.}
\label{fig:1PD}
\end{figure}

To the best of our knowledge, no phase diagram for the system Nd$_2$O$_3$--Lu$_2$O$_3$ is published so far. Hence, the model of Nd$_2$O$_3$--Lu$_2$O$_3$ in Fig.~\ref{fig:1PD} is derived from phase diagrams of comparable systems with similar radii of RE$^{3+}$ ions and is adapted to phase compositions and transitions that were reported recently for Nd$_2$O$_3$:Lu$_2$O$_3$=1:1 by the current authors \cite{Hirsch17}, and completed by new data. The ``comparable systems'' that were used are Nd$_2$O$_3$--Y$_2$O$_3$ and Nd$_2$O$_3$--Yb$_2$O$_3$ \cite{Coutures74,Tikhonov77,Adylov87}; combined with a section through the system Pr$_2$O$_3$--Lu$_2$O$_3$ at 1673\,K \cite{Berndt76}. Recently it was shown that annealed mixtures with the nominal composition NdLuO$_3$ consist of a mixture of C and B up to at least 1923\,K (diamonds in Fig.~\ref{fig:1PD}, \cite{Hirsch17}). Phase transition temperatures measured by DTA for the ``NdLuO$_3$'' composition that were identified in that publication, together with recent data are shown in Fig.~\ref{fig:1PD} as empty diamonds. It is remarkable that the P-phase does not appear in the Nd$_2$O$_3$--Lu$_2$O$_3$ system, and that the H- and X-phase show wide homogeneity ranges.

\begin{figure}[ht]
\includegraphics[width=0.60\textwidth]{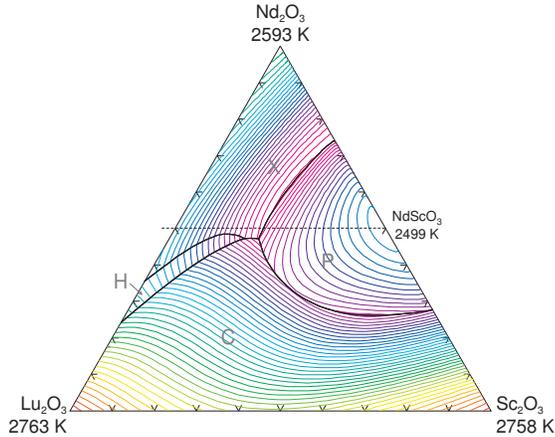}
\caption{Liquidus surface in the system Nd$_2$O$_3$--Lu$_2$O$_3$--Sc$_2$O$_3$ with 10\,K isotherms. Congruent melting points as labeled.}
\label{fig:liquidus}
\end{figure}

The definition of the pseudoternary system includes all thermodynamic values and functions of the three pseudobinary systems in Fig.~\ref{fig:2PDs} and \ref{fig:1PD}, together with the ternary interaction function (\ref{eq:Gex-ter}) and binary interaction between P-NdLuO$_3$ and P-NdScO$_3$. From this data set, the projection onto the liquidus surface of the concentration triangle in Fig.~\ref{fig:liquidus} was calculated. The four regions of primary crystallization are marked by appropriate labels.

\begin{figure}[ht]
\includegraphics[width=0.60\textwidth]{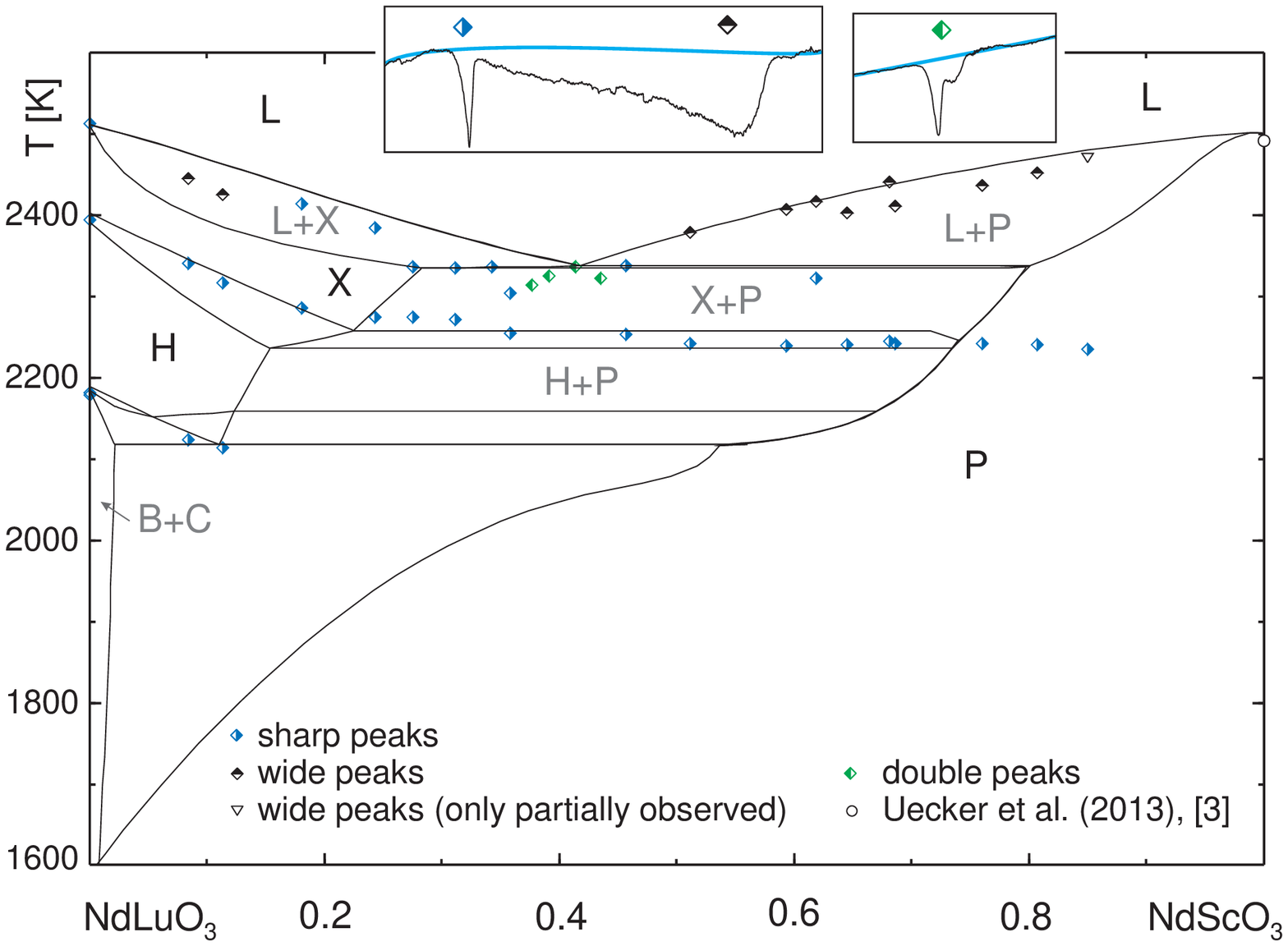}
\caption{Assessed isopleth section NdLuO$_3$--NdScO$_3$, together with DTA peaks and with the melting point of P-NdScO$_3$ \cite{Uecker13}. The accuracy of DTA derived temperatures is approximately $\pm$15\,K for all data. Different symbols are used for different peak shapes (cf. inlays).}
\label{fig:NdLuO$_3$-NdScO$_3$-DTA}
\end{figure}

As expected, the P-phase field extends widely around the congruently melting NdScO$_3$ compound; this means that the growth of perovskite mixed crystals should be possible from melt compositions ranging almost to the middle of the concentration triangle. The thermodynamic data that allowed the calculation of Fig.~\ref{fig:liquidus} resulted in a congruent melting point of 2499\,K for NdScO$_3$, which is in good agreement with the recent experimental value of 2491\,K \cite{Uecker13}. From Fig.~\ref{fig:liquidus} the composition of the crystallizing solid phase(s) cannot be read. The thermodynamic data set allows the calculation of isopleth $x-T$ sections through the triangle, and in Fig.~\ref{fig:NdLuO$_3$-NdScO$_3$-DTA} this is performed for the ``NdLuO$_3$''--NdScO$_3$ section that corresponds to the dashed line in Fig.~\ref{fig:liquidus}. As an overlay, Fig.~\ref{fig:NdLuO$_3$-NdScO$_3$-DTA} shows the experimental points that were obtained by DTA measurements on this section.

The agreement between experiment and calculation is reasonable, especially for very high $T$ close to the liquidus. Larger discrepancies, and especially missing experimental signals in the subsolidus region are common, because transformation heats between solid phases are usually smaller than the heat for melting. Besides, the transformation of solid phases requires often diffusion steps that are kinetically hindered, which further reduces thermal effects that could be measured by DTA. Segregation inside molten DTA samples can cause other issues such as the upper boundary of the H+P field that extends with unlikely points to the P-phase field. But if a melt ``L'' e.g. with composition $x=0.8$ solidifies, a P-phase with $x\approx0.94$ crystallizes first. This shifts the melt composition in Fig.~\ref{fig:NdLuO$_3$-NdScO$_3$-DTA} downwards the liquidus to the direction of the eutectic. In the $x$ range near the eutectic the X-phase can be formed, which later on converts to H, giving rise to the observation of the three rightmost DTA points inside the P-phase field. The wide stability of the P-phase at lower $T$ is in agreement with the X-ray data from the Figs.~\ref{fig:x-ray} and \ref{fig:Vegard}.

\section{Crystal growth}

\begin{figure}[ht]
\includegraphics[width=0.60\textwidth]{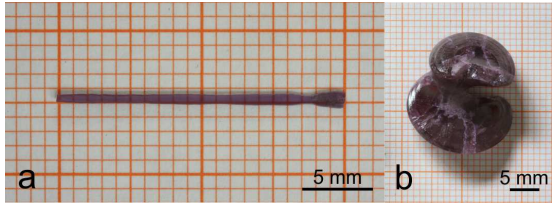}
\caption{a) $\mu$-PD fibre grown from a starting composition NdLu$_{0.15}$Sc$_{0.85}$O$_3$. b) Czochralski boule with strong spiraling, grown from a starting composition NdLu$_{0.40}$Sc$_{0.60}$O$_3$.}
\label{fig:foto}
\end{figure}

The possibility of crystal growth within the P-phase field that contains the congruently melting NdScO$_3$ compound is demonstrated by the micro-pulling-down ($\mu$-PD) crystal in Fig.~\ref{fig:foto}; this polished section along the length axis evidences the internal quality. Furthermore, Czochralski growth was performed to receive larger samples for further investigations (see below). The multicrystalline boule exhibits the shape of a spiral, most likely caused by a leaky crucible. Never\-the\-less, it was good enough for examination with XRD and ICP-OES. The X-ray diffractogram revealed a pattern typical for orthorhombic rare-earth scandates, with lattice parameters $a=5.813\pm0.005$\,\AA, $b=8.058\pm0.005$\,\AA, $c=5.593\pm0.005$\,\AA\ for sample T2 in Table~\ref{tab:crystals}. From these data a pseudocubic lattice constant $a_\mathrm{pc}=4.031$\,\AA\ can be calculated using the relation $a_\mathrm{pc}=\sqrt[3]{\frac{1}{4}abc}$ \cite{Ubic07}. Chemical data for the crystals are summarized in Tab.~\ref{tab:crystals}.

Obviously, the crystal is depleted in Nd with respect to the melt and its composition does not lie on the isopleth NdLuO$_3$--NdScO$_3$. The measured chemical compostion of the crystal shows a significant offset to the composition that can be expected from the liquidus/solidus positions in Fig.~\ref{fig:NdLuO$_3$-NdScO$_3$-DTA}. This indicates, that under the applied growth conditions the effective partition coefficients $k_\mathrm{eff}$ are shifted from the equilibrium partition coefficients $k_0$ towards unity, which is common for $k_\mathrm{eff}$ \cite{Burton53}. For the following reasons it is expected, that the crystal chemistry of melt-grown NdLu$_{1-x}$Sc$_x$O$_3$ solid solutions is rather complex: First, vacancies on the larger cations side could replace Nd. Such vacancies are described for several rare-earth scandates \cite{Uecker08}. Second, a partial replacement of Nd by Lu seems to be likely. Replacement of La on the large cation site was proven for LaLuO$_3$ \cite{Ovanesyan99} but also in the case of LaLu$_{1-x}$Sc$_x$O$_3$ single crystals \cite{Uecker17}. Perhaps, this substitution by Lu is stronger in a crystal with Nd on the large cation site than for La because the difference in ionic radii between Nd and Lu is smaller than for La and Lu.

\begin{table}[ht]
\caption{Chemical composition as determined by ICP-OES of original melts for crystal growth with micro-pulling-down ($\mu$-PD) and Czochralski (Cz) methods, and pieces of the Cz crystal. T5 represents the first crystallizate. T2 and T4 show the further development.}
\begin{tabular}{lccc}
\hline 
Sample                     & Nd$_2$O$_3$ [mol-\%] &  Lu$_2$O$_3$ [mol-\%] & Sc$_2$O$_3$ [mol-\%]  \\
\hline
Original melt $\mu$-PD     & 49.3                 & 7.6                   & 43.1   \\
Original melt Cz           & 50.0                 & 20.0                  & 30.0   \\
NdLu$_{1-x}$Sc$_x$O$_3$ T5 & 45.6                 & 14.0                  & 40.6   \\
NdLu$_{1-x}$Sc$_x$O$_3$ T2 & 45.5                 & 14.7                  & 39.8   \\
NdLu$_{1-x}$Sc$_x$O$_3$ T4 & 45.8                 & 15.2                  & 39.1   \\
\hline
\end{tabular}
\label{tab:crystals}
\end{table}

\section{Conclusions}

A model of the pseudoternary system Nd$_2$O$_3$--Lu$_2$O$_3$--Sc$_2$O$_3$ was obtained by DTA and XRD measurements and thermodynamic assessment. The thermodynamic parameter set models the pseudobinary edge systems very well, and also experimental results that were measured of the NdLuO$_3$--NdScO$_3$ isopleth section can be understood well in terms of the pseudoternary model. Because DTA measurements were the main experimental method of this study, basically, processes with significant thermal effects contributed to the results --- and these are mainly melting/solidification processes. Accordingly, Fig.~\ref{fig:liquidus}, showing the liquidus projection with four regions of primary crystallization of different phases is assumed to be accurate. The isopleth section in Fig.~\ref{fig:NdLuO$_3$-NdScO$_3$-DTA} must be handled with more care: certainly the diagram is correct in the high-$T$ region above approximately 2200\,K because sufficient experimental data is available there. At lower $T$, however, DTA effects practically did not occur because solid state phase transformations proceed slowly and hence thermal effects are small. This means that the phase boundaries that are computed there are mainly the result of extrapolation of high-$T$ results to this range. A few artifacts of the calculation, e.g. a miscibility gap of the P-phase at $T\leq1800$\,K are not shown in Fig.~\ref{fig:NdLuO$_3$-NdScO$_3$-DTA} because no experimental evidence could be found. But also here, too low diffusivity could be the reason. The wide homogeneity range of the P-phase at moderate $T\lesssim2000$\,K is remarkable and could path the way for the production of ceramic materials.

The main goal of this paper was to reveal conditions where the growth of P-NdLu$_{1-x}$Sc$_x$O$_3$ mixed crystals might be possible. Fig.~\ref{fig:liquidus} shows that from melts with $0.45\lesssim x\leq1$ the P-phase crystallizes first. From Fig.~\ref{fig:NdLuO$_3$-NdScO$_3$-DTA} it can be seen that even from melts close to the eutectic valley on the isopleth ($x\approx0.45$) crystals with considerably reduced NdLuO$_3$ content can be obtained as a result of strong segregation. This is confirmed by a Czochralski crystal growth experiment on this section, where a mixed crystal with pseudocubic lattice constant $a_\mathrm{pc}=4.031$\,\AA\ was obtained.

\section*{Acknowledgments}

The authors express their gratitude to Rainer Bertram and Mario Br\"utzam (both IKZ Berlin) for performing chemical analysis by ICP-OES and for help with Czochralski growth experiments, respectively. Prof. Jens G\"otze (TU Bergakademie Freiberg) is thanked for supervising the first author's MA thesis.

\section*{References}


\begin{thebibliography}{10}

\bibitem{Schlom14}
D.~G. Schlom, L.-Q. Chen, C.~J. Fennie, V.~Gopalan, D.~A. Muller, X.~Pan, R.~Ramesh, R.~Uecker, Elastic strain engineering of ferroic oxides, MRS Bull.
 39 (2014) 118--130.

\bibitem{Uecker08}
R.~Uecker, B.~Velickov, D.~Klimm, R.~Bertram, M.~Bernhagen, M.~Rabe, M.~Albrecht, R.~Fornari, D.~G. Schlom, Properties of rare-earth scandate
 single crystals {(RE=Nd--Dy)}, J. Crystal Growth 310 (2008) 2649--2658.

\bibitem{Uecker13}
R.~Uecker, D.~Klimm, R.~Bertram, M.~Bernhagen, I.~Schulze-Jonack, M.~Br{\"{u}}tzam, A.~Kwasniewski, T.~Gesing, D.~G. Schlom, Growth and investigation of {Nd$_{1-x}$Sm$_x$ScO$_3$} and {Sm$_{1-x}$Gd$_x$ScO$_3$} solid-solution single crystals, Acta Physica Polonica A 124 (2013) 295--300.

\bibitem{Berndt76}
U.~Berndt, D.~Maier, C.~Keller, {Phasengleichgewichte in Inter\-lanthaniden\-oxid-Systemen}, J. Solid State Chem. 16 (1976) 189--195, in German.

\bibitem{Uecker17}
R.~Uecker, R.~Bertram, M.~Br{\"{u}}tzam, Z.~Galazka, T.~M. Gesing, C.~Guguschev, D.~Klimm, M.~Klupsch, A.~Kwasniewski, D.~G. Schlom, Large-lattice-parameter perovskite single-crystal substrates, J. Crystal Growth 457 (2017) 137--142.

\bibitem{Sayed15}
F.~N. Sayed, R.~Shukla, A.~K. Tyagi, A chemical method for stabilizing a new series of solid solution {Pr$_{1-x}$Ce$_x$ScO$_3$} $(0.0\leq x\leq1.0)$
  systems, Dalton Trans. 44 (2015) 16929--16936.

\bibitem{Guguschev18}
C.~Guguschev, J.~Hidde, T.~M. Gesing, M.~Gogolin, D.~Klimm, {Czochralski} growth and characterization of {Tb$_x$Gd$_{1-x}$ScO$_3$} and
{Tb$_x$Dy$_{1-x}$ScO$_3$} solid-solution single crystals, CrystEngComm (2018) in the press.

\bibitem{Coutures76}
J.~Coutures, J.~P. Coutures, Etude des solutions solides ``{CeLnO$_3$}'' et des perovskites {CeLnO$_3$} et {PrLnO$_3$} ({Ln} = \'el\'ement lanthanidique), J. Solid State Chem. 19 (1976) 29--33, in French.

\bibitem{Coutures80}
J.-P. Coutures, J.~M. Badie, R.~Berjoan, J.~Coutures, R.~Flamand, A.~Rouanet, Stability and thermodynamic properties of rare earth perovskites, High
 temperature science 13 (1980) 331--336.

\bibitem{Badie78}
J.-M. Badie, Phases et transitions de phases {\`{a}} haute temp{\'{e}}rature dans les syst{\`{e}}mes {Sc$_2$O$_3$--Ln$_2$O$_3$} ({Ln} = lanthanide et
  yttrium), Revue Internationale des Hautes Temp\'eratures et des R\'efractaires 15 (1978) 183--199, in French.

\bibitem{Gesing09}
T.~M. Gesing, R.~Uecker, J.~Christian-Buhl, Refinement of the crystal structure of praseodymium orthoscandate, {PrScO$_3$}, Z. Krist. -- New Crystal
  Structures 224 (2009) 385--386.

\bibitem{Porotnikov80a}
N.~V. Porotnikov, O.~I. Kondratov, K.~I. Petrov, I.~I. Olikov, Vibrational spectra of double rare-earth oxides with composition {LnLuO$_3$}, Zh. Neorg.
 Chim. 25 (1980) 699--704, in Russian.

\bibitem{Porotnikov83}
N.~V. Porotnikov, V.~N. Cygankov, N.~B. Gorilovskaya, K.~I. Petrov, About some double rare-earth oxides with rhombic perovskite structure, Zh. Neorg. Chim.
 28 (1983) 2799--2805, in Russian.

\bibitem{Coutures74}
J.~Coutures, R.~Verges, M.~Foex, Etude a haute temperature des systemes formes par le sesquioxyde de neodyme avec les sesquioxydes d'yttrium et d'ytterbium, Mat. Res. Bull. 9 (1974) 1603--1612, in French.

\bibitem{Berndt75}
U.~Berndt, D.~Maier, C.~Keller, New {A$^\mathrm{III}$B$^\mathrm{III}$O$_3$} interlanthanide perovskite compounds, J. Solid State Chem. 13 (1975)
  131--135.

\bibitem{Schneider60a}
S.~J. Schneider, R.~S. Roth, Phase equilibria in systems involving the rare-earth oxides. {Part II. Solid} state reactions in trivalent rare-earth
 oxide systems, J. Res. Nat. Bureau Stand. Sect. A: Physics and Chemistry 64A (1960) 317--332.

\bibitem{Bharathy09}
M.~Bharathy, A.~H. Fox, S.~J. Mugavero, H.-C. zur Loye, Crystal growth of inter-lanthanide {LaLn$'$O$_3$ (Ln$'$=Y, Ho--Lu}) perovskites from hydroxide
  fluxes, Solid State Sciences 11 (2009) 651--654.

\bibitem{Ito01}
K.~Ito, K.~Tezuka, Y.~Hinatsu, Preparation, magnetic susceptibility, and specific heat on interlanthanide perovskites {ABO$_3$ (A=La--Nd, B=Dy--Lu)},
 J. Solid State Chem. 157 (2001) 173--179.

\bibitem{Hirsch17}
T.~Hirsch, R.~Uecker, D.~Klimm, Reevaluation of phase relations in the chemical system neodymium lutetium oxide {NdLuO$_3$}, Cryst. Res. Technol. 52 (2017)
  1600237.

\bibitem{Konings14}
R.~J.~M. Konings, O.~Bene{\v{s}}, A.~Kov{\'{a}}cs, D.~Manara, D.~Sedmidubsk{\'{y}}, L.~Gorokhov, V.~S. Iorish, V.~Yungman, E.~Shenyavskaya,
  E.~Osina, The thermodynamic properties of the $f$-elements and their compounds. $part$ 2. the lanthanide and actinide oxides, J. Phys. Chem. Ref.
 Data 43 (2014) 1--95.

\bibitem{Adachi98}
G.~Adachi, N.~Imanaka, The binary rare earth oxides, Chemical Reviews 98 (1998) 1479--1514.

\bibitem{Coutures76a}
J.~Coutures, F.~Sibieude, M.~Foex, Etude a haute temp{\'{e}}rature des syst{\`{e}}mes form{\'{e}}s par les sesquioxydes de lanthane avec les
 sesquioxydes de lanthanides {II}. {Influence} de la trempe sur la nature des phases obtenues {\`{a}} la temp{\'{e}}rature ambiante, J. Solid State Chem.
  17 (1976) 377--384, in French.

\bibitem{Badie70}
J.-M. Badie, {\'{E}}tude de la structure des phases {\`{a}} haute temp{\'{e}}rature present{\'{e}}es pas les syst{\`{e}}mes
 {Sc$_2$O$_3$--La$_2$O$_3$} et {Sc$_2$O$_3$--Nd$_2$O$_3$}, High Temperatures -- High Pressures 2 (1970) 309--316, in French.

\bibitem{Burns89}
G.~W. Burns, M.~G. Scroger, The calibration of thermocouples and thermocouple materials, Nat. Inst. Stand. Technol. (1989) 25.

\bibitem{FactSage7_1}
{www.factsage.com}, FactSage 7.1, GTT Technologies, Kaiserstr. 100, 52134 Herzogenrath, Germany (2017).

\bibitem{Wojdyr10}
M.~Wojdyr, {{\it Fityk}: a general-purpose peak fitting program}, J. Appl. Cryst. 43 (2010) 1126--1128.

\bibitem{Shannon76}
R.~D. Shannon, Revised effective ionic radii and systematic studies of interatomic distances in halides and chalcogenides, Acta Cryst. A 32 (1976)
  751--767.

\bibitem{Vegard21}
L.~Vegard, {Die Konstitution der Mischkristalle und die Raumf\"ullung der Atome}, Z. Phys. 5 (1921) 17--26, in German.

\bibitem{Uecker15}
R.~Uecker, New large-lattice-constant perovskite single-crystal substrates, in: {Leibniz-Institut f\"ur Kristallz\"uchtung} (Ed.), Annual Report,
 {www.ikz-berlin.de}, 2015, pp. 38--47.

\bibitem{Zinkevich07}
M.~Zinkevich, Thermodynamics of rare earth sesquioxides, Prog. Mat. Sci. 52 (2007) 597--647.

\bibitem{Qi15}
J.~Qi, X.~Guo, A.~Mielewczyk-Gryn, A.~Navrotsky, Formation enthalpies of {LaLnO$_3$ (Ln=Ho, Er, Tm and Yb)} interlanthanide perovskites, J. Solid
  State Chem. 227 (2015) 150--154.

\bibitem{Klimm15}
D.~Klimm, {Phase Equilibria}, in: T.~Nishinaga (Ed.), Handbook of Crystal Growth (Second Edition), 2nd Edition, Elsevier, Amsterdam, 2015, pp. 85--136.

\bibitem{Redlich48}
O.~Redlich, A.~T. Kister, Algebraic representation of thermodynamic properties and the classification of solutions, Industrial {\&} Engineering Chemistry 40 (1948) 345--348.

\bibitem{Tikhonov77}
P.~A. Tikhonov, A.~K. Kuznetsov, E.~F. Zhikhareva, K.~Y. Merezhinskii, V.~N. Yuneev, Phase diagrams {Y$_2$O$_3$--Nd$_2$O$_3$} and physico-chemical
 properties of solid solutions, Zh. Neorg. Chim. 22 (1977) 1057--1061, in Russian.

\bibitem{Adylov87}
G.~T. Adylov, G.~V. Voronov, L.~M. Sigalov, The system {Nd$_2$O$_3$--Y$_2$O$_3$}, Inorg. Mat. 23 (1987) 1867--1870, in Russian.

\bibitem{Ubic07}
R.~Ubic, Revised method for the prediction of lattice constants in cubic and pseudocubic perovskites, J. Amer. Ceram. Soc. 90~(10) (2007) 3326--3330.

\bibitem{Burton53}
J.~A. Burton, R.~C. Prim, W.~P. Slichter, Distribution of solute in crystals grown from the melt, J. Chem. Phys. 21 (1953) 1987--1991.

\bibitem{Ovanesyan99}
K.~L. Ovanesyan, A.~G. Petrosyan, G.~O. Shirinyan, C.~Pedrini, L.~Zhang, {Czochralski} single crystal growth of {Ce-} and {Pr-}doped {LaLuO$_3$}
  double oxide, J. Crystal Growth 198-199 (1999) 497--500.

\end{thebibliography}

\end{document}